\documentclass[12pt]{article}
\usepackage{epsfig,amsmath,amssymb,mathrsfs}
\usepackage[utf8]{inputenc}
\usepackage[colorlinks=true,citecolor=blue,,linktocpage=true,linkcolor=blue,urlcolor=black]{hyperref}
\usepackage{mathtools,slashed}
\usepackage{array}
\usepackage{xcolor}
\usepackage[force]{feynmp-auto}
\usepackage{caption}
\usepackage{subcaption}

\usepackage[left=2.5cm,bottom=3cm,right=2.5cm,top=3cm]{geometry}

\usepackage{overpic,youngtab}
\usepackage[latin,english]{babel}
\usepackage{amsmath}
\usepackage{amssymb}
\usepackage{epstopdf}
\usepackage{graphics,psfrag,rotating}
\usepackage{mathabx}
\usepackage{graphicx}
\usepackage{dcolumn}
\usepackage{float}
\usepackage{pdflscape}
\usepackage{array}
\usepackage{booktabs}
\usepackage{amscd}
\usepackage{mathtools}
\usepackage{fancybox}
\usepackage{tensor}
\usepackage{fix-cm}
\usepackage[colorlinks=true,citecolor=blue,,linktocpage=true,linkcolor=blue,urlcolor=black]{hyperref}
\usepackage{cleveref}
\usepackage{tikz}
\usetikzlibrary{calc}
\usetikzlibrary{intersections}
\usetikzlibrary{positioning}
\usetikzlibrary{arrows}
\usetikzlibrary{shapes.misc}

\usetikzlibrary{patterns,snakes}
\tikzstyle{spring}=[line width=0.8,black,snake=coil,segment amplitude=4.25,segment length=4.75,line cap=round]
\usetikzlibrary{decorations.pathmorphing}
\usetikzlibrary{decorations.markings}
\usetikzlibrary{arrows.meta}
\usetikzlibrary{matrix}
\usetikzlibrary{positioning}
\usetikzlibrary{arrows}
\usepackage{titlesec}
\usepackage{abstract}
\usepackage{cancel}
\usepackage{mathtools}
\usepackage{pdflscape}
\usepackage{graphicx}
\definecolor{mypink}{HTML}{FDA4BA}
\usepackage{pdflscape}
\usepackage{subcaption}

\usepackage{dcolumn}

\usepackage{dashrule}        
\usepackage[compat=1.1.0]{tikz-feynman}   

  {\end{list}}%

\makeatletter
\renewcommand{\section}{\setcounter{equation}{0}\@startsection
 {section}%
 {1}%
 {0pt}%
 {-1\baselineskip}%
 {0.4\baselineskip}%
 {\bfseries\large}}%
\renewcommand{\subsection}{\@startsection
 {subsection}%
 {2}%
 {0pt}%
 {-0.75\baselineskip}%
 {0.2\baselineskip}%
 {\bfseries}}%
\renewcommand{\subsubsection}{\@startsection
 {subsubsection}%
 {3}%
 {0pt}%
 {-0.5\baselineskip}%
 {0.1\baselineskip}%
 {\sc}}%
\makeatother

\DeclareMathAlphabet{\mathpzc}{OT1}{pzc}{m}{it}

\def\be{\begin{equation}}
\def\ee{\end{equation}}

\def\g5{\gamma_{5}}

\def\id3k{\int\!\! \dfrac{d^3\!\vec{k}}{(2\pi)^3 2E(\vec{k})}}

\newcommand{\bea}{\begin{eqnarray}}
\newcommand{\eea}{\end{eqnarray}}
\newcommand{\beann}{\begin{eqnarray*}}
\newcommand{\eeann}{\end{eqnarray*}}
\newcommand{\ba}{\begin{array}}
\newcommand{\ea}{\end{array}}

 \def\g {\gamma}

\newcommand{\email}[1]{\href{mailto:#1}{\tt #1}}

\begin{document}

\rightline{\scriptsize{IPARCOS-UCM-23-113}}
\vglue 50pt

\begin{center}

{\LARGE \bf Quantization of Weyl invariant unimodular gravity with antisymmetric ghost fields}\\
\vskip 1.0true cm
{\Large David Garc\'{\i}a-L\'opez$^{\dagger}$, Carmelo P. Martin$^{\dagger\dagger}$}
\\
\vskip .7cm
{
	{Universidad Complutense de Madrid (UCM), Departamento de Física Teórica and IPARCOS, Facultad de Ciencias Físicas, 28040 Madrid, Spain}
	
	\vskip .5cm
	\begin{minipage}[l]{.9\textwidth}
		\begin{center}
			\textit{E-mail:}
			\email{$\dagger$davgar20@ucm.es},
			\email{$\dagger\dagger$carmelop@fis.ucm.es}.

		\end{center}
	\end{minipage}
}
\end{center}
\thispagestyle{empty}

\begin{abstract}
The enforcement of the unimodularity condition in a gravity theory by means of a Lagrange multiplier  leads, in general, to inconsistencies upon quantization. This is so, in particular, when the classic linear splitting of the metric between the background and quantum fields is used. To avoid the need of introducing such a Lagrange multiplier while using the classic linear splitting, we  carry out the quantization of unimodular gravity with extra Weyl symmetry by using Becchi-Rouet-Stora-Tyutin (BRST) techniques. Here, two gauge symmetries are to be gauge-fixed: transverse diffeomorphisms and Weyl transformations. We perform the gauge-fixing of the transverse diffeomorphism invariance by using  BRST transformations that involve antisymmetric ghost fields. We show that these BRST transformations are compatible with the BRST transformations needed to gauge-fix the Weyl symmetry, so that they can be combined in a set of transformations generated by a single BRST operator. Newton's law of gravitation is derived within the BRST formalism we put forward as well as the Slavnov-Taylor equation.
\end{abstract}

\vfill
\clearpage

\section{Introduction.}

Unimodular gravity --see Refs.~\cite{Carballo-Rubio:2022ofy} and \cite{Alvarez:2023utn}, for introductions-- solves \cite{vanderBij:1981ym, Zee:1983jg, Buchmuller:1988wx, Henneaux:1989zc, Buchmuller:2022msj} a part of the so-called Cosmological Constant problem, for in that gravity theory the vacuum energy does not   gravitate. Indeed, all the field configurations in unimodular gravity have got determinant equal to $-1$.

There exists sundry approaches to the quantization of unimodular gravity, e.g., Refs.~\cite{Eichhorn:2013xr, Padilla:2014yea, Alvarez:2015sba, Bufalo:2015wda,  deLeonArdon:2017qzg, DeBrito:2019gdd, Baulieu:2020obv, deBrito:2021pmw, Kugo:2022iob}. It is not known whether they give rise to the same quantum theory on a Minkowski background, for they involve different sets of ghosts. This is an open problem, as it is their equivalence to quantum General Relativity when the Cosmological Constant is not set to zero.

It has been shown recently --see Ref.~\cite{Alvarez:2023kuw}-- that  the unimodularity condition cannot be implemented in the path integral by using a Lagrange multiplier unless the unimodularity condition be equivalent to imposing a linear constraint on the graviton field. This fact precludes the use of the classic linear splitting of the metric into background metric and graviton field if the unimodularity condition is to be enforced by using a Lagrange multiplier. As also shown in
Ref.~\cite{Alvarez:2023kuw} the problem goes away if the so-called exponential splitting \cite{Ohta:2016npm} is used.  We think that it is quite a drawback to have a formalism where the classic linear splitting --which is standardly used in General Relativity-- cannot be employed. This is so even sidestepping the issue that even in General Relativity the exponential parametrization may not yield the same quantum theory as the linear splitting: the exponential parametrization is not a   field redefinition of the linear splitting -see Ref.~\cite{Nink:2016dpi} and Refs. therein.

 In Refs.~\cite{Padilla:2014yea, Bufalo:2015wda, Baulieu:2020obv} and \cite{Kugo:2022iob} the unimodularity condition is imposed by using a Lagrange multiplier and thus those formalisms do not make sense for the linear splitting of the metric. In Refs.~\cite{Eichhorn:2013xr,  deLeonArdon:2017qzg, DeBrito:2019gdd} and \cite{deBrito:2021pmw} the exponential parametrization is implemented already at classical level and prior to quantization. The BRST formalism put forward in Ref.~\cite{Alvarez:2015sba} admits any kind of parametrization in terms of background and quantum fields  --the linear and exponential parametrizations, in particular-- for no Lagrange multiplier is introduced. Indeed, in Ref.~\cite{Alvarez:2015sba}, building on Refs.~\cite{Alvarez:2006uu} and \cite{Alvarez:2005iy}, the unimodularity condition is solved, prior to quantization, by expressing the unimodular metric, $\hat{g}_{\mu\nu}$, in terms of an unconstrained tensor field $g_{\mu\nu}$ as follows
\begin{equation}
\hat{g}_{\mu\nu}(x)=\dfrac{g_{\mu\nu}(x)}{|g|^{1/D}(x)}.
\label{unconstrainedg}
\end{equation}
$D$ is the space-time dimension. Several properties of the formalism set up in Ref.~\cite{Alvarez:2015sba} have been analyzed leading to sound results 
--see Refs.~\cite{Anero:2022rqi}, \cite{Garay:2023nco}  and \cite{Anero:2023xjb}, for recent publications. Let us recall that because of Eq.~(\ref{unconstrainedg})    the classical action of unimodular gravity has two gauge symmetries: transverse diffeomorphisms and Weyl transformations of $g_{\mu\nu}$. This is why the gravity theory  in question is called Weyl invariant unimodular gravity.

As we have pointed out above, the BRST formalism put forward in Ref.~\cite{Kugo:2022iob} breaksdown at the quantum level if the classic linear splitting of the metric is employed. This is the parametrization originally  considered in Ref.~\cite{Kugo:2022iob}. Of course, the inconsistency disappears if the exponential parametrization is used. The question arises as to whether this formalism can be modified so that the classic linear splitting can be  utilized consistently.

The purpose of this paper is to show that, by using Eq.~(\ref{unconstrainedg}), the framework of Ref.~\cite{Kugo:2022iob} can be combined with part of the formalism of Ref.~\cite{Alvarez:2015sba} so that the linear splitting of the gravity field works at the quantum level: no Lagrange multiplier is needed here to enforce the unimodularity condition. That this combination works is non trivial since the implementation of Eq.~(\ref{unconstrainedg}) introduces an additional gauge symmetry --a Weyl symmetry-- in the theory. This gauge symmetry leads to additional BRST transformations which might clash with BRST transformations arising from the transverse gauge transformations as defined in Ref.~\cite{Kugo:2022iob}.

The layout of this paper is as follows. In section 2 we discuss the Becchi-Rouet-Stora-Tyutin (BRST) quantization of our unimodular theory. This second section contains several subsections since our unimodular theory has two gauge symmetries, namely, transverse diffeomorphisms and Weyl transformations. In subsection 2.1, we carry out the gauge-fixing of the invariance under transverse diffeomorhisms by adapting to our case the BRST formalism of  Ref.~\cite{Kugo:2022iob}. Subsection 2.2 is devoted to the analysis of the gauge-fixing of the  Weyl symmetry by using BRST methods. In this latter subsection, we show the compatibility of the BRST transformations associated to transverse diffeomorphism and the BRST transformations coming from the Weyl transformations. We prove that the two BRST operators, each generating one of those two BRST transformations, can be combined into a single BRST operator. This single operator gives rise to the full BRST invariance of the theory. In subsection 2.3, we give the full BRST invariant action of the theory. We make a consistency check of the BRST formalism we have put forward by obtaining, in section 3, Newton's potential. In section 4, we derive the Slavnov-Taylor equation, and the corresponding linearized Slavnov-Taylor operator, for cohomological techniques will be needed to obtain sensible renormalized radiative corrections: as we shall discuss, dimensional regularization does not preserve the BRST symmetry of the theory.

\section{BRST quantization.}

In 4 space-time dimensions, the classical action of the unimodular theory with Weyl invariance reads
\begin{equation}
    \mathcal{S}_{class}=2\int d^4x\ R\left[\hat{g}\right],
    \label{clasaction}
\end{equation}
where $\hat{g}_{\mu\nu}$ is defined  in terms of the unconstrained field $g_{\mu\nu}$ in Eq.~(\ref{unconstrainedg}), $g_{\mu\nu}$ being the dynamical field variable. $R\left[\hat{g}\right]$ above denotes the Ricci scalar for unimodular metric $\hat{g}_{\mu\nu}$.

The  action in Eq.~(\ref{clasaction}) has two gauge symmetries acting on $g_{\mu\nu}$: it is invariant under transverse diffeomorphisms and under
Weyl transformations. The infinitesimal transverse diffeomorphisms $x^\mu\to x^\mu+\xi_\text{T}^\mu$, with $\partial_\mu \xi_\text{T}^\mu=0$,  are defined as follows:
\begin{equation}
g_{\mu\nu}\to g_{\mu\nu}+\delta g_{\mu\nu},\quad
\delta g_{\mu\nu}=\nabla_{\mu}\xi^\text{T}_{\nu}+\nabla_{\nu}\xi^\text{T}_{\mu},
\label{transdiff}
\end{equation}
where $\xi^\text{T}_\mu\equiv g_{\mu\nu}\xi_\text{T}^\nu$ and $\nabla_\mu$ is the covariant derivative for $g_{\mu\nu}$. The infinitesimal Weyl transformations read
\begin{equation*}
g_{\mu\nu}\to g_{\mu\nu}+\delta g_{\mu\nu},\quad
\delta g_{\mu\nu}=2\theta g_{\mu\nu},
\end{equation*}
$\theta$ being an infinitesimal scalar field. Notice that these transformations leave $\hat{g}_{\mu\nu}$, as defined in Eq.~(\ref{unconstrainedg}), invariant by construction.

Before going ahead, let us show that the transformation in (\ref{transdiff}) generates an infinitesimal transverse diffeomorphism of $\hat{g}_{\mu\nu}$:
\begin{equation}
\begin{array}{l}
   {\delta\hat{g}_{\mu\nu}= \delta\left(\frac{g_{\mu\nu}}{(-g)^{1/4}}\right)=\frac{1}{(-g)^{1/4}}\left(\delta g_{\mu\nu}-\frac{1}{4g}g_{\mu\nu}\delta g\right)}\\[8pt]
    {\phantom{\delta\hat{g}_{\mu\nu}}=\frac{1}{(-g)^{1/4}}\left(\nabla_{\mu}\xi^\text{T}_{\nu}+\nabla_{\nu}\xi^\text{T}_{\mu}-\frac{1}{2}\nabla_{\mu}\xi_\text{T}^{\mu}\right)
    =\hat{\nabla}_{\mu}\hat{\xi}^\text{T}_{\nu}+\hat{\nabla}_{\nu}\hat{\xi}^\text{T}_{\mu},}
\end{array}
\label{proofofgeneration}
    \end{equation}
where $\hat{\xi}^\text{T}_\mu=\hat{g}_{\mu\nu}\xi_\text{T}^\nu$ and $\hat{\nabla}_\mu$ is the covariant derivative with regard to $\hat{g}_{\mu\nu}$. The previous variation of $\hat{g}_{\mu\nu}$ leaves its determinant invariant, since $\hat{\nabla}_{\mu}\xi_\text{T}^{\mu}=\partial_\mu\xi_\text{T}^{\mu}$ for a unimodular metric.
Indeed
\begin{equation}
    \delta\hat{g}=\hat{g}\hat{g}^{\mu\nu}\delta \hat{g}_{\mu\nu}
    =-\hat{g}^{\mu\nu} \left(\hat{\nabla}_{\mu}\hat{\xi}^\text{T}_{\nu}+\hat{\nabla}_{\nu}\hat{\xi}^\text{T}_{\mu}\right)
    =-2\hat{\nabla}_{\mu}\xi_\text{T}^{\mu}
    =-2\partial_{\mu}\xi_\text{T}^{\mu}=0.
    \label{zerovariation}
\end{equation}

Below we shall use
\begin{equation}
 \delta g^{\mu\nu}=-\nabla^{\mu}\xi_\text{T}^{\nu}-\nabla^{\nu}\xi_\text{T}^{\mu}, \quad
 \delta\hat{ g}^{\mu\nu}=-\hat{\nabla}^{\mu}\xi_\text{T}^{\nu}-\hat{\nabla}^{\nu}\xi_\text{T}^{\mu},
 \label{alternative}
 \end{equation}
which are equivalent to Eqs. (\ref{transdiff}) and (\ref{proofofgeneration}), respectively

\subsection{BRST formalism for the transverse diffeomorphisms.}

In this subsection we shall adapt the formalism put forward in Ref.~\cite{Kugo:2022iob} to case at hand, i.e., the metric, $\hat{g}_{\mu\nu}$, will be unimodular from the very beginning, its transverse diffeomorphisms being generated by the transverse diffeomorphisms of $g_{\mu\nu}$ --see Eqs. (\ref{transdiff}),  (\ref{proofofgeneration}) and (\ref{zerovariation})-- and the transversality of $\xi_\text{T}^{\mu}$ is not with regard to the covariant derivative of $g_{\mu\nu}$, but    with regard to the partial derivative. This entails a  modification, though minor, of the formalism in Ref.~\cite{Kugo:2022iob}, for  transversality in  Ref.~\cite{Kugo:2022iob} --see its Eq.~(1.8)-- is with regard to the covariant derivative of the metric; this metric being  unimodular only after having solved for the Lagrange multiplier.

Although only a minor modification of the formalism in Ref.~\cite{Kugo:2022iob} is needed for our transverse diffeomorphisms, we shall give all the relevant details to make this subsection self-consistent and to set the notation.

Let $\mathfrak{B}_D$ denote the BRST operator with ghost number 1 and let us introduce the transverse ghost fields $c_\text{T}^\mu$ ($N_\text{ghost}(c)=1$) and $\overline{c}_\text{T}^\nu$ ($N_\text{ghost}(\overline{c})=-1$), where $N_\text{ghost}$ stands for the ghost number. The action of $\mathfrak{B}_D$  on $g^{\mu\nu}$   reads
\begin{equation}
\mathfrak{B}_D g^{\mu\nu}=-\nabla^\mu c_\text{T}^\nu-\nabla^\nu c_\text{T}^\mu,
\label{brsgmunu}
\end{equation}
which in turn --see Eq.~(\ref{proofofgeneration})-- induces the following BRST transformation of $\hat{g}^{\mu\nu}$:
\begin{equation}
     \mathfrak{B}_D \hat{g}^{\mu\nu}=-\hat{\nabla}^\mu c_\text{T}^\nu-\hat{\nabla}^\nu c_\text{T}^\mu.
\label{brshatg}
\end{equation}
Eqs.~(\ref{brsgmunu}) and (\ref{brshatg}) are the BRST counterparts of the transformations in Eq.~(\ref{alternative}), respectively. Note that $\nabla_\mu$ denotes the covariant derivative with regard to $g_{\mu\nu}$.

We should keep in mind that when a linear splitting of $g_{\mu\nu}$ into background, $\overline{g}_{\mu\nu}$, and quantum, $h_{\mu\nu}$, fields is carried out,
\begin{equation*}
g_{\mu\nu}=\overline{g}_{\mu\nu}+h_{\mu\nu},
\end{equation*}
the BRST transformation in Eq.~(\ref{brsgmunu}) boils down, by definition, to
\begin{equation}
\begin{array}{l}
{\mathfrak{B}_D \overline{g}_{\mu\nu}=0,}\\[4pt]
{\mathfrak{B}_D h_{\mu\nu}=\mathfrak{B}_D g_{\mu\nu}=\nabla_\mu c^\text{T}_\nu+\nabla_\nu c^\text{T}_\mu=}\\[4pt]
{\quad\quad
\overline{\nabla}_\mu(\overline{g}_{\nu\rho}c_\text{T}^\rho)+\overline{\nabla}_\nu(\overline{g}_{\mu\rho}c_\text{T}^\rho)+
c_\text{T}^\rho\overline{\nabla}_\rho h_{\mu\nu}+(\overline{\nabla}_\mu c_\text{T}^\rho) h_{\rho\nu}+(\overline{\nabla}_\nu c_\text{T}^\rho) h_{\rho\mu}.}
\end{array}
\label{BRSTforhmn}
\end{equation}
$\overline{\nabla}_\mu$ is the covariant derivative with regard to $\overline{g}_{\mu\nu}$.

The action of the BRST operator on the ghost fields is defined to be
\begin{equation}
    \mathfrak{B}_D c_\text{T}^\mu=\hat{\nabla}_\nu(c_\text{T}^\nu c_\text{T}^\mu),
 \label{BcT}
\end{equation}
\begin{equation}
    \mathfrak{B}_D \overline{c}_\text{T}^\mu=ib_\text{T}^\mu,
  \label{BbarcT}
\end{equation}
\begin{equation}
    \mathfrak{B}_D b_\text{T}^\mu=0.
    \label{BbT}
\end{equation}
 $b_\text{T}^\mu$ is the transverse \textit{Nakanishi-Lautrup} auxiliary field. Let us recall that
 \begin{equation}
 \hat{\nabla}_{\mu}c_\text{T}^\mu=\partial_{\mu}c_\text{T}^\mu=0,\quad \hat{\nabla}_{\mu}\overline{c}_\text{T}^\mu=\partial_{\mu}\overline{c}_\text{T}^\mu=0\quad\text{and that}\quad\hat{\nabla}
 _{\mu}b_\text{T}^\mu=\partial_{\mu}b_\text{T}^\mu=0.
 \label{transversality}
 \end{equation}
 The reader should always bear in mind that, for any vector field $V^\mu$,
 \begin{equation*}
 \hat{\nabla}_\mu V^\mu = \partial_\mu V^\mu,
 \end{equation*}
 since $\hat{g}_{\mu\nu}$ has determinant equal to $-1$.

 Taking into account the previous definitions and that
 \begin{equation}
 \hat{\nabla}_\nu(c_\text{T}^\nu c_\text{T}^\mu)=c_\text{T}^\nu\hat{\nabla}_\nu c_\text{T}^\mu=c_\text{T}^\nu\partial_\nu c_\text{T}^\mu,
 \label{usefuleq}
 \end{equation}
 it is easy to show that  $\mathfrak{B}_D$ is nilpotent:
 \begin{equation*}
 \mathfrak{B}_D^2=0.
 \end{equation*}

Once we have defined the action of the BRST operator on the fields, we may try to use, in a na\"{\i}ve way, the BRST quantization procedure and add to the classical lagrangian the following BRST-exact object
\begin{equation}
    \mathcal{L}_\text{ghost}^{(D)}=-i\mathfrak{B}_D\left[\hat{g}_{\mu\nu}\overline{c}_\text{T}^\nu\partial_\lambda \hat{g}^{\lambda\mu}\right],
\label{Lagrangianone}
\end{equation}
to define the path integral. But this will not do for the ghost fields we have introduced are constrained fields and we do not know its integration measure. To overcome this difficulty we shall solve the constraints as done in Ref.~\cite{Kugo:2022iob}.  Indeed, the constraints in Eq.~(\ref{transversality}) can be solved by introducing three unconstrained, antisymmetric tensor fields $c^{\nu\mu}$, $\overline{c}^{\nu\mu}$ and $b^{\nu\mu}$ and express $c_\text{T}^{\mu}$, $\overline{c}_\text{T}^{\mu}$ and $b_\text{T}^{\mu}$ in terms of the former as follows
\begin{equation}
    c_\text{T}^\mu=\hat{\nabla}_\nu c^{\nu\mu},\quad \overline{c}_\text{T}^\mu=\hat{\nabla}_\nu\overline{c}^{\nu\mu},\quad\text{and}\quad b_\text{T}^\mu=\hat{\nabla}_\nu b^{\nu\mu}.
\label{solvingtheconstraint}
\end{equation}
Notice that  $\hat{\nabla}_\nu c^{\nu\mu}=\partial_\nu c^{\nu\mu}$, and so on, since the metric $\hat{g}_{\mu\nu}$ is unimodular; however, we have decided to keep the  covariant derivative for $\hat{g}_{\mu\nu}$ to render any comparison with the results in Ref.~\cite{Kugo:2022iob} straightforward.

We may now define the path integral of the theory by integrating over de unconstrained fields $c^{\nu\mu}$, $\overline{c}^{\nu\mu}$ and $b^{\nu\mu}$. But, unfortunately that path integral will not be well defined because the use of Eq.~(\ref{solvingtheconstraint}) will give rise to a new gauge invariance: the substitutions
\begin{equation}
c^{\nu\mu}\to c^{\nu\mu}+\hat{\nabla}_\rho d^{\rho\nu\mu},\quad \overline{c}^{\nu\mu}\to \overline{d}^{\nu\mu}+\hat{\nabla}_\rho \overline{d}^{\rho\nu\mu},\quad
b^{\nu\mu}\to b^{\nu\mu}+\hat{\nabla}_\rho b^{\rho\nu\mu}
\label{newgaugesymone}
\end{equation}
do not change the values of $c_\text{T}^\mu$, $\overline{c}_\text{T}^\mu$ and $b_\text{T}^\mu$, respectively, if  $c^{\rho\nu\mu}$, $\overline{d}^{\rho\nu\mu}$ and $b^{\rho\nu\mu}$ are antisymmetric tensor fields. This gauge invariance comes from the mismatch between the 3 independent components of, say, $c_\text{T}^\mu$ and the 6 independent components of $c^{\rho\nu\mu}$. We shall fix this mismatch, and remove the corresponding gauge invariance, by gauge-fixing. This gauge-fixing  will  demand the introduction of new ghosts --called second generation ghosts, which will be antisymmetric tensor fields of rank 3. This second generation ghosts will give rise in turn to a new gauge symmetry, which will lead to the introduction of a third generation of new ghosts --rank 4 antisymmetric tensors.
This process is called ghost for ghost process. As we shall see below the ghost for ghost process stops after introducing the rank 4 tensor ghosts if the dimension of space-time is 4, but this is not so if the dimension of space-time is greater than 4. We thus conclude that the field content of the quantum theory depends on the spacetime dimension.

\subsubsection{BRST variations of the first generation ghosts.}

Let us work out the action of the BRST operator $\mathfrak{B}_D$ on the first generation ghosts $c^{\nu\mu}$, $\overline{c}^{\nu\mu}$ and $b^{\nu\mu}$, which are the rank 2 antisymmetric tensor fields introduced in Eq.~(\ref{solvingtheconstraint}).

Let $V^{\mu\nu_1...\nu_n}$ a rank $n+1$ antisymmetric tensor field. Then the following result
\begin{equation*}
    \hat{\nabla}_\mu V^{\mu\nu_1...\nu_n}=\partial_\mu\left(V^{\mu\nu_1...\nu_n}\right)
\end{equation*}
leads to
\begin{equation*}
    \mathfrak{B}_D\left(\hat{\nabla}_\mu V^{\mu\nu_1...\nu_n}\right)=\hat{\nabla}_\mu\left(\mathfrak{B}_D V^{\mu\nu_1...\nu_n}\right).
\end{equation*}

Taking into account the previous equation and  Eqs.~(\ref{BcT}) and (\ref{solvingtheconstraint}), one gets
\begin{equation*}
    \hat{\nabla}_\nu\left(\mathfrak{B}_D c^{\nu\mu}-c_\text{T}^\nu c_\text{T}^\mu\right)=0.
\end{equation*}
The general solution to this equation reads
\begin{equation*}
    \mathfrak{B}_D c^{\nu\mu}=c_\text{T}^\nu c_\text{T}^\mu+i\hat{\nabla}_\rho d^{\rho\nu\mu}.
\end{equation*}
The symbol $d^{\rho\nu\mu}$ denotes a new ghost which is an antisymmetric tensor field of  rank 3 and it is called a second generation ghost.

We are now ready to define the action of $\mathfrak{B}_D$ on the first generation ghost, antighost and auxiliary field:
\begin{equation}
\begin{array}{l}
    {\mathfrak{B}_D c^{\nu\mu}=c_\text{T}^\nu c_\text{T}^\mu+i\hat{\nabla}_\rho d^{\rho\nu\mu},}\\[8pt]
    {\mathfrak{B}_D \overline{c}^{\nu\mu}=ib^{\nu\mu},}\\[8pt]
    {\mathfrak{B}_D b^{\nu\mu}=0}.
\end{array}
\label{Bc2Bb2}
\end{equation}
respectively. Notice that the two last equations above are consistent with Eqs.~(\ref{BbarcT}) and (\ref{BbT}), respectively.

Let us express $\mathcal{L}_\text{ghost}^{(D)}$ in Eq.~(\ref{Lagrangianone}) in terms of the first generation ghosts:
\begin{equation}
\begin{array}{l}
    {\mathcal{L}_\text{ghost,1}^{(D)}\equiv \mathcal{L}_\text{ghost}^{(D)}=
    -i\mathfrak{B}_D\left[\hat{g}_{\mu\nu}\hat{\nabla}_\rho\overline{c}^{\rho\nu}\cdot\partial_\lambda \hat{g}^{\lambda\mu}\right]}\\[8pt]
{\phantom{\mathcal{L}_\text{ghost,1}^{(D)}}=\hat{g}_{\mu\nu}\hat{\nabla}_\rho b^{\rho\nu}\cdot\partial_\lambda \hat{g}^{\lambda\mu}
+2i\hat{\nabla}_{(\mu} \hat{\nabla}^\rho c_{\nu)\rho}\cdot\hat{\nabla}_\rho\overline{c}^{\rho\nu}\cdot\partial_\lambda \hat{g}^{\lambda\mu}
+2i\hat{g}_{\mu\nu}\hat{\nabla}_\rho\overline{c}^{\rho\mu}\cdot\partial_\lambda\hat{\nabla}^{(\lambda} \hat{\nabla}_\rho c^{\mu)\rho}.}
\end{array}
\label{Lagrangian1}
\end{equation}
The symbol $(\mu_1,\mu_2...)$ stands for full  symmetrization so that already symmetric objects are left invariant.

\subsubsection{Second generation ghosts and BRST transformations.}

We have already discussed in the last paragraph of subsection 2.1 that the transformations in Eq.~(\ref{newgaugesymone}) are gauge symmetries of the theory, the fields $d^{\rho\nu\mu}$, $\overline{d}^{\rho\nu\mu}$ and $b^{\rho\nu\mu}$  being referred to as second generation ghosts. It is apparent that the ghost Lagrangian in Eq.~(\ref{Lagrangian1}) is invariant under the changes in   Eq.~(\ref{newgaugesymone}). Notice that in the previous subsection the field $d^{\rho\nu\mu}$ has shown up  already in the BRST transformations of the first generation ghost $c^{\nu\mu}$ --see Eq.~(\ref{Bc2Bb2}).

To remove the gauge invariance generated by the need of  second generation ghosts one introduces gauge-fixing conditions. The gauge-fixing conditions we choose read
\begin{equation}
\nabla^{[\rho}c^{\nu\mu]}=0\quad\text{and}\quad    \nabla^{[\rho}\overline{c}^{\nu\mu]}=0.
\label{GFIX}
\end{equation}
The  Lagrange multiplier BRST  doublets needed to implement them in the Lagrangian  are
\begin{equation}
(\overline{d}_{\rho\nu\mu},\overline{c}_{\rho\nu\mu})\quad\text{and}\quad(b_{\rho\nu\mu},c_{\rho\nu\mu}),
\label{firstdoublets}
\end{equation}
respectively. The  symbol $[\mu_1,\mu_2...]$ denotes antisymmetrization with regard to the indices so that antisymmetric indices are left invariant. Note that the gauge-fixing term for the gauge transformation of $b^{\nu\mu}$ is obtained using the BRST transformation of $\overline{c}^{\nu\mu}$.

Let us now introduce the BRST transformations  of the second generation ghosts. The first equation in (\ref{Bc2Bb2}) and the requirement that $\mathfrak{B}_D^2=0$ leads to the following equation to be satisfied by $\mathfrak{B}_D d^{\rho\nu\mu}$:
\begin{equation*}
    \hat{\nabla}_\rho\left(i\mathfrak{B}_D d^{\rho\nu\mu}+c_\text{T}^\rho c_\text{T}^\nu c_\text{T}^\mu\right)=0.
\end{equation*}
The general solution to this equation runs thus:
\begin{equation}
    \mathfrak{B}_D d^{\rho\nu\mu}=ic_\text{T}^\rho c_\text{T}^\nu c_\text{T}^\mu-\hat{\nabla}_\sigma t^{\sigma\rho\nu\mu}.
    \label{BRSTofd3}
\end{equation}
The symbol $t^{\sigma\rho\nu\mu}$ stands for a new ghost field which is a rank 4 antisymmetric field, i.e., a third generation ghost.

We are now ready to define the action of the BRST operator on the second generation ghosts:
\begin{equation}
\begin{array}{l}
    {\mathfrak{B}_D d^{\rho\nu\mu}=ic_\text{T}^\rho c_\text{T}^\nu c_\text{T}^\mu-\hat{\nabla}_\sigma t^{\sigma\rho\nu\mu},}\\[4pt]
    {\mathfrak{B}_D \overline{d}^{\rho\nu\mu}=\overline{c}^{\rho\nu\mu},}\\[4pt]
    {\mathfrak{B}_D b^{\rho\nu\mu}=c^{\rho\nu\mu},}\\[4pt]
    {\mathfrak{B}_D c^{\rho\nu\mu}=\mathfrak{B}_D \overline{c}^{\rho\nu\mu}=0.}
\end{array}
 \label{BdBb3}
\end{equation}
The new fields  $\overline{c}^{\rho\nu\mu}$ and $c^{\rho\nu\mu}$ constitute BRST doublets with $\overline{d}^{\rho\nu\mu}$ and $ b^{\rho\nu\mu}$, respectively. Recall
Eq.~(\ref{firstdoublets}) and the key point that indices are raised and lowered with the unimodular metric $\hat{g}_{\mu\nu}$.

Furnished with the gauge-fixing conditions in Eq.~(\ref{GFIX}) and the BRST transformations in Eq.~(\ref{BdBb3}), we are ready to introduce a ghost Lagrangian that removes the gauge symmetry arising from the introduction of second generation ghosts, i.e., the gauge invariance under the transformations in  Eq.~(\ref{newgaugesymone}):
\begin{equation}
\begin{array}{l}
    {\mathcal{L}_\text{ghost,2}^{(D)}=\frac{i}{2}\mathfrak{B}_D\left[\overline{d}^{\rho\nu\mu}\hat{\nabla}_\rho c_{\nu\mu}+b^{\rho\nu\mu}\hat{\nabla}_\rho \overline{c}_{\nu\mu}\right]}\\[8pt]
    {=\frac{i}{2}\left[\overline{c}^{\rho\nu\mu}\hat{\nabla}_\rho c_{\nu\mu}-\hat{\nabla}_\rho\overline{d}^{\rho\nu\mu} \cdot\mathfrak{B}_D\left(\hat{g}_{\nu\sigma}\hat{g}_{\mu\kappa}\right)c^{\sigma\kappa}\right.
    -\hat{\nabla}^\rho\overline{d}_{\rho\nu\mu}\cdot\left(c_\text{T}^\nu c_\text{T}^\mu+i\hat{\nabla}_\sigma d^{\sigma\nu\mu}\right)
    + c^{\rho\nu\mu}\hat{\nabla}_\rho \overline{c}_{\nu\mu}}\\[8pt]
    {\quad\quad\;-\hat{\nabla}_\rho b^{\rho\nu\mu}\cdot\mathfrak{B}_D\left(\hat{g}_{\nu\sigma}\hat{g}_{\mu\kappa}\right)\overline{c}^{\sigma\kappa}
    \left.-i\hat{\nabla}^\rho b_{\rho\nu\mu}\cdot b^{\nu\mu}\right].}
\end{array}
\label{Lagrangian2}
\end{equation}

\subsubsection{Third generation ghosts and their BRST transformations.}

At least one gauge symmetry still occurs in our theory --we have not discussed the Weyl symmetry yet. This gauge symmetry is generated by rank 4 antisymmetric fields and arises from the invariance of the ghost Lagrangian in Eq.~(\ref{Lagrangian2}) under the following transformations
\begin{equation}
\begin{array}{l}
    {d^{\rho\nu\mu}\to d^{\rho\nu\mu}-\hat{\nabla}_\sigma t^{\sigma\rho\nu\mu},}\\[4pt]
     {\overline{d}^{\rho\nu\mu}\to \overline{d}^{\rho\nu\mu}+\hat{\nabla}_\sigma\overline{t}^{\sigma\rho\nu\mu},\quad b^{\rho\nu\mu}\to b^{\rho\nu\mu}+\hat{\nabla}_\sigma b^{\sigma\rho\nu\mu},}\\[4pt]
    {c^{\rho\nu\mu}\to c^{\rho\nu\mu}+\hat{\nabla}_\sigma c^{\sigma\rho\nu\mu},\quad \overline{b}^{\rho\nu\mu}\to \overline{c}^{\rho\nu\mu}+\hat{\nabla}_\sigma \overline{c}^{\sigma\rho\nu\mu}.}
    \end{array}
   \label{zeta3zeta4}
\end{equation}
The fields $t^{\sigma\rho\nu\mu}$, $\overline{t}^{\sigma\rho\nu\mu}$, $b^{\sigma\rho\nu\mu}$, $\overline{b}^{\sigma\rho\nu\mu}$, $c^{\sigma\rho\nu\mu}$ and $\overline{c}^{\sigma\rho\nu\mu}$, which generate the previous gauge transformations will be promoted to ghost fields becoming third generation ghosts.

To remove the gauge invariance of the Lagrangian under the gauge transformations in Eq.~(\ref{zeta3zeta4}), we shall introduce --as done in Ref.~\cite{Kugo:2022iob}--
the following gauge-fixing conditions:
\begin{equation}
    \nabla^{[\sigma}d^{\rho\nu\mu]}=0,\quad
    \nabla^{[\sigma}\overline{d}^{\rho\nu\mu]}=0\quad\text{and}\quad
     \nabla^{[\sigma}b^{\rho\nu\mu]}=0.
 \label{gf3}
\end{equation}
The Lagrange multiplier BRST  doublets used to implement the previous gauge-fixing conditions are the following
\begin{equation*}
(\overline{t}_{\sigma\rho\nu\mu}, \overline{d}_{\sigma\rho\nu\mu}),\quad
(c_{\sigma\rho\nu\mu}, d_{\sigma\rho\nu\mu})\quad\text{and}\quad (\overline{c}_{\sigma\rho\nu\mu}, b_{\sigma\rho\nu\mu}),
\end{equation*}
respectively.

In the previous list, we have not included any gauge-fixing conditions for $c^{\rho\nu\mu}$ and $\overline{c}^{\rho\nu\mu}$ since they will be generated, via the action of the BRST operator,  by the gauge-fixing terms in the Lagrangian that implement the gauge-fixing conditions for $b^{\rho\nu\mu}$ and $\overline{d}^{\rho\nu\mu}$, respectively --see Eq.~(\ref{Lagrangian3}) below.

Let us define next the BRST transformations for the third generation ghosts $t^{\sigma\rho\nu\mu}$, $\overline{t}^{\sigma\rho\nu\mu}$, $b^{\sigma\rho\nu\mu}$, $c^{\sigma\rho\nu\mu}$ and $\overline{c}^{\sigma\rho\nu\mu}$. Let us begin with the BRST transformation of $t^{\sigma\rho\nu\mu}$. Since we must have $\mathfrak{B}_D^2=0$, Eq.~(\ref{BRSTofd3}) leads to
\begin{equation*}
    \hat{\nabla}_\sigma\left(\mathfrak{B}_D t^{\sigma\rho\nu\mu}-ic_\text{T}^\sigma c_\text{T}^\rho c_\text{T}^\nu c_\text{T}^\mu\right)=0.
\end{equation*}
The general solution to this equation reads
\begin{equation*}
    \mathfrak{B}_D t^{\sigma\rho\nu\mu}=ic_\text{T}^\sigma c_\text{T}^\rho c_\text{T}^\nu c_\text{T}^\mu.
\end{equation*}
No new ghosts occur this time since, in 4 dimensions, no rank 5 antisymmetric tensor exists. We are now in the right position to define the BRST  transformations
\begin{equation}
\begin{array}{l}
    {\mathfrak{B}_D t^{\sigma\rho\nu\mu}=ic_\text{T}^\sigma c_\text{T}^\rho c_\text{T}^\nu c_\text{T}^\mu,\quad
    \mathfrak{B}_D \overline{t}^{\sigma\rho\nu\mu}=i\overline{d}^{\sigma\rho\nu\mu},}\\[4pt]
    {\mathfrak{B}_D c^{\sigma\rho\nu\mu}=id^{\sigma\rho\nu\mu},\quad
    \mathfrak{B}_D \overline{c}^{\sigma\rho\nu\mu}=ib^{\sigma\rho\nu\mu},}\\[4pt]
    {\mathfrak{B}_D \overline{d}^{\sigma\rho\nu\mu}=\mathfrak{B}_D d^{\sigma\rho\nu\mu}=\mathfrak{B}_D b^{\sigma\rho\nu\mu}=0.}
    \end{array}
    \label{fourgenBRST}
\end{equation}

The gauge-fixing conditions in Eq.~(\ref{gf3}) and the BRST transformations in Eq.~(\ref{fourgenBRST}) lead to the following part of the BRST invariant Lagrangian which carries out the gauge-fixing of the symmetries in Eq.~(\ref{zeta3zeta4}):
\begin{equation}
\begin{array}{l}
    {\mathcal{L}_\text{ghost,3}^{(D)}=\frac{-i}{6}\mathfrak{B}_D\left[-\overline{t}^{\sigma\rho\nu\mu}\left(\hat{\nabla}_\sigma d_{\rho\nu\mu}+\frac{\alpha}{4}d_{\sigma\rho\nu\mu}\right)+c^{\sigma\rho\nu\mu}\hat{\nabla}_\sigma \overline{d}_{\rho\nu\mu}+\overline{c}^{\sigma\rho\nu\mu}\hat{\nabla}_\sigma b_{\rho\nu\mu}\right]}\\[8pt]
    {\phantom{\mathcal{L}_\text{ghost,3}^{(D)}}=\frac{1}{6} \Big[-\overline{d}^{\sigma\rho\nu\mu}\hat{\nabla}_\sigma d_{\rho\nu\mu}-\frac{\alpha}{4}\overline{d}^{\sigma\rho\nu\mu}d_{\sigma\rho\nu\mu} +
    i\hat{\nabla}^\sigma\overline{t}_{\sigma\rho\nu\mu}\cdot\left(i c_\text{T}^\rho c_\text{T}^\nu c_\text{T}^\mu-\hat{\nabla}_\lambda t^{\lambda\rho\nu\mu}\right)}\\[8pt]
    {\phantom{\mathcal{L}_\text{ghost,3}^{(D)}=\frac{1}{6}[}
    +d^{\sigma\rho\nu\mu}\hat{\nabla}_\sigma \overline{d}_{\rho\nu\mu}+i c^{\sigma\rho\nu\mu}\hat{\nabla}_\sigma \overline{c}_{\rho\nu\mu}
    +b^{\sigma\rho\nu\mu}\hat{\nabla}_\sigma b_{\rho\nu\mu}+i\overline{c}^{\sigma\rho\nu\mu}\hat{\nabla}_\sigma c_{\rho\nu\mu}}\\[8pt]
    {\phantom{\mathcal{L}_\text{ghost,3}^{(D)}=\frac{1}{6}[}
    +i\hat{\nabla}_\sigma\overline{t}^{\sigma\rho\nu\mu} \cdot\mathfrak{B}_D\left(\hat{g}_{\rho\kappa}\hat{g}_{\nu\tau}\hat{g}_{\mu\lambda}\right)d^{\kappa\tau\lambda}-i\hat{\nabla}_\sigma c^{\sigma\rho\nu\mu} \cdot\mathfrak{B}_D\left(\hat{g}_{\rho\kappa}\hat{g}_{\nu\tau}\hat{g}_{\mu\lambda}\right)\overline{d}^{\kappa\tau\lambda}}\\[8pt]
    {\phantom{\mathcal{L}_\text{ghost,3}^{(D)}=\frac{1}{6}[}
    -i\hat{\nabla}_\sigma\overline{c}^{\sigma\rho\nu\mu} \cdot\mathfrak{B}_D\left(\hat{g}_{\rho\kappa}\hat{g}_{\nu\tau}\hat{g}_{\mu\lambda}\right)b^{\kappa\tau\lambda}\Big].}
\end{array}
\label{Lagrangian3}
\end{equation}

\subsubsection{The ghost gauge-fixing Lagrangian for the transverse diffeomorphisms.}

The gauge invariance of the theory under transverse diffeomorphisms is fully gauge-fixed in a BRST invariant way by considering a ghost Lagrangian, $\mathcal{L}_\text{ghost}^{(D)}$, which is the sum of $\mathcal{L}_\text{ghost,1}^{(D)}$, $\mathcal{L}_\text{ghost,2}^{(D)}$ and $\mathcal{L}_\text{ghost,3}^{(D)}$ given in Eqs.~(\ref{Lagrangian1}), (\ref{Lagrangian2}) and (\ref{Lagrangian3}), respectively.

Next, following Ref.~\cite{Kugo:2022iob}, we shall express $\mathcal{L}_\text{ghost}^{(D)}$ in terms of the Hodge duals of ghosts fields: Hodge duals with regard to the unimodular metric $g_{\mu\nu}$. Let us now defined these Hodge duals:
\begin{eqnarray*}
    c^{\mu\nu}&=&(1/2)\varepsilon^{\mu\nu\rho\sigma}C_{\rho\sigma},\nonumber\\
    d^{\mu\nu\rho}&=&\varepsilon^{\mu\nu\rho\sigma}D_{\sigma},\nonumber\\
    t^{\mu\nu\rho\sigma}&=&\varepsilon^{\mu\nu\rho\sigma}T.
\end{eqnarray*}
and
\begin{eqnarray*}
    \left(\begin{matrix}
    \overline{c}^{\mu\nu}\\
    b^{\mu\nu}
    \end{matrix}\right) & = & -(1/2)\varepsilon^{\mu\nu\rho\sigma}
    \left(\begin{matrix}
    \overline{C}_{\rho\sigma}\\
    B_{\rho\sigma}
    \end{matrix}\right),\hspace{1cm}
    \left(\begin{matrix}
    \overline{t}^{\mu\nu\rho\sigma}\\
    \overline{d}^{\mu\nu\rho\sigma}
    \end{matrix}\right) = -\varepsilon^{\mu\nu\rho\sigma}
    \left(\begin{matrix}
    \overline{T}\\
    \overline{D}
    \end{matrix}\right),\nonumber\\
    \left(\begin{matrix}
    \overline{d}^{\mu\nu\rho}\\
    \overline{c}^{\mu\nu\rho}
    \end{matrix}\right) & = & -(1/2)\varepsilon^{\mu\nu\rho\sigma}
    \left(\begin{matrix}
    \overline{D}_{\sigma}\\
    \overline{C}_{\sigma}
    \end{matrix}\right),\hspace{1cm}
    \left(\begin{matrix}
    c^{\mu\nu\rho\sigma}\\
    d^{\mu\nu\rho\sigma}
    \end{matrix}\right) = -\varepsilon^{\mu\nu\rho\sigma}
    \left(\begin{matrix}
    C\\
    D
    \end{matrix}\right),\nonumber\\
    \left(\begin{matrix}
    b^{\mu\nu\rho}\\
    c^{\mu\nu\rho}
    \end{matrix}\right) & = & -(1/2)\varepsilon^{\mu\nu\rho\sigma}
    \left(\begin{matrix}
    B_{\sigma}\\
    C_{\sigma}
    \end{matrix}\right),\hspace{1cm}
    \left(\begin{matrix}
    \overline{c}^{\mu\nu\rho\sigma}\\
    b^{\mu\nu\rho\sigma}
    \end{matrix}\right) = -\varepsilon^{\mu\nu\rho\sigma}
    \left(\begin{matrix}
    \overline{C}\\
    B
    \end{matrix}\right).
\end{eqnarray*}

Hence, we get
\begin{equation}
\mathcal{L}_\text{ghost}^{(D)}=\mathcal{L}_\text{ghost,1}^{(D)}+\mathcal{L}_\text{ghost,2}^{(D)}+\mathcal{L}_\text{ghost,3}^{(D)},
\label{translagrangian}
\end{equation}
where
\begin{equation}
\begin{array}{l}
    {\mathcal{L}_\text{ghost,1}^{(D)}=-\frac{1}{2}\hat{g}_{\mu\nu}\varepsilon^{\rho\nu\sigma\tau}\hat{\nabla}_\rho B^{\sigma\tau}\cdot\partial_\lambda \hat{g}^{\lambda\mu}-\frac{i}{2}\varepsilon^{\chi\tau\rho\sigma}\varepsilon^{\nu\kappa\alpha\beta}\hat{g}_{\chi(\mu}\hat{\nabla}_{\nu)}\hat{\nabla}_\tau C_{\rho\sigma}\cdot\hat{\nabla}_{\kappa}\overline{C}_{\alpha\beta}\cdot\partial_\lambda \hat{g}^{\lambda\mu}}\\[4pt]
{+\frac{i}{2}\hat{g}_{\mu\nu}\varepsilon^{\nu\rho\sigma\tau}\hat{\nabla}_\rho \overline{C}^{\sigma\tau}\cdot\varepsilon^{\kappa\alpha\beta(\mu}\partial_\lambda\hat{\nabla}^{\lambda)}\hat{\nabla}_{\kappa}C_{\alpha\beta},}
\end{array}
\label{Lagra1}
\end{equation}
\begin{equation}
\begin{array}{l}
    {\mathcal{L}_\text{ghost,2}^{(D)}=i\overline{C}^{\sigma}\hat{\nabla}^\rho C_{\rho\sigma} +\frac{3i}{4}\varepsilon^{\kappa\mu\nu\lambda} \hat{\nabla}^\rho\overline{D}^\sigma\cdot\hat{\nabla}_{[\rho}C_{\kappa\sigma]}\cdot\hat{\nabla}_\mu C_{\nu\lambda}+2\hat{\nabla}^{[\lambda}\overline{D}^{\sigma]}\cdot\hat{\nabla}_\lambda D_\sigma}\\[4pt] {\quad\quad-iC^\sigma\hat{\nabla}^\rho\overline{C}_{\rho\sigma}+B^\sigma\hat{\nabla}^\rho B_{\rho\sigma}
    -2i\hat{g}^{\mu\lambda}\hat{\nabla}^{[\rho}\overline{D}^{\sigma]}\cdot C_{\mu\sigma}\mathfrak{B}_D\hat{g}_{\lambda\rho}+
    2i\hat{g}^{\mu\lambda}\hat{\nabla}^{[\rho}B^{\sigma]}\cdot\overline{C}_{\mu\sigma}\mathfrak{B}_D\hat{g}_{\lambda\rho},}
\end{array}
\label{Lagra2}
\end{equation}
and
\begin{equation}
\begin{array}{l}
    {\mathcal{L}_\text{ghost,3}^{(D)}=\Big[\overline{D}\hat{\nabla}_\mu D^{\mu} +\alpha \overline{D}D+D\hat{\nabla}_\mu\overline{D}^\mu +iC\hat{\nabla}_\mu \overline{C}^\mu +i\overline{C}\hat{\nabla}_\mu C^\mu +B\hat{\nabla}_\mu B^\mu}\\[4pt]
    {\quad\quad-\frac{1}{4}\hat{\nabla}^\sigma\overline{T}\cdot\hat{\nabla}_{[\sigma}C_{\nu\mu]}\big(\hat{\nabla}^\mu C^{\lambda\rho}\cdot\hat{\nabla}^\nu C_{\lambda\rho} -4\hat{\nabla}^\lambda C^{\mu\rho}\cdot\hat{\nabla}^\nu C_{\lambda\rho} +2\hat{\nabla}^\lambda C^{\mu\rho}\cdot\hat{\nabla}_\lambda {C^\nu}_\rho }\\[4pt]
    {\quad\quad -2\hat{\nabla}_\lambda C^{\mu\rho}\cdot\hat{\nabla}_\rho C^{\nu\lambda}\big)
    -i\hat{\nabla}^\mu\overline{T}\cdot\hat{\nabla}_\mu T+\left(i\hat{\nabla}^\mu\overline{T}\cdot D^\nu+i\hat{\nabla}^\mu C\cdot\overline{D}^\nu+i\hat{\nabla}^\mu\overline{C}\cdot B^\nu\right)\mathfrak{B}_D\hat{g}_{\mu\nu}\Big].}
\end{array}
\label{Lagra3}
\end{equation}
$\mathcal{L}_\text{ghost}^{(D)}$ above can be obtained from the corresponding expression in Ref.~\cite{Kugo:2022iob} by choosing in the latter the unimodular metric $\hat{g}_{\mu\nu}$ as the spacetime metric.

In table 1, the reader may find a summary of the action of $\mathfrak{B}_D$ on the Hodge duals introduced above

\begin{table}[h]
    \hspace{-3mm}
    \scalebox{0.8}{
    \begin{tabular}{m{2.2cm}lll}
        & 1st generation & 2nd generation & 3rd generation\\\hline
        & & & \\
         $N_{ghost}=0$ & $\mathfrak{B}_D B_{\mu\nu}=0$ & $\mathfrak{B}_D B_{\mu}=C_{\mu}$ & $\mathfrak{B}_D B=0$  \\
         & & & \\
         $N_{ghost}=1$ & $\mathfrak{B}_D C_{\mu\nu}=-\frac{1}{2}\varepsilon_{\mu\nu\rho\sigma}c_\text{T}^\rho c_\text{T}^\sigma+i\left(\partial_\mu D_\nu-\partial_\nu D_\mu\right)$ & $\mathfrak{B}_D C_{\mu}=0$ & $\mathfrak{B}_D C=iD$  \\
         & & & \\
         $N_{ghost}=-1$ & $\mathfrak{B}_D\overline{C}_{\mu\nu}=iB_{\mu\nu}$ & $\mathfrak{B}_D \overline{C}_{\mu}=0$ & $\mathfrak{B}_D \overline{C}=iB$  \\
         & & & \\
         $N_{ghost}=2$ & \hspace{1cm}\hdashrule[3pt]{1cm}{0.5pt}{1mm} & $\mathfrak{B}_D D_{\mu}=\frac{i}{3!}\varepsilon_{\mu\nu\rho\sigma}c_\text{T}^\nu c_\text{T}^\rho c_\text{T}^\sigma +\partial_\mu T$ & $\mathfrak{B}_D D=0$  \\
         & & & \\
         $N_{ghost}=-2$ & \hspace{1cm}\hdashrule[3pt]{1cm}{0.5pt}{1mm} & $\mathfrak{B}_D \overline{D}_{\mu}=\overline{C}_{\mu}$ & $\mathfrak{B}_D \overline{D}=0$  \\
         & & & \\
         $N_{ghost}=3$ & \hspace{1cm}\hdashrule[3pt]{1cm}{0.5pt}{1mm} & \hspace{1cm}\hdashrule[3pt]{1cm}{0.5pt}{1mm} & $\mathfrak{B}_D T=-\frac{i}{4!}\varepsilon_{\mu\nu\rho\sigma}c_\text{T}^\mu c_\text{T}^\nu c_\text{T}^\rho c_\text{T}^\sigma$  \\
         & & & \\
         $N_{ghost}=-3$ & \hspace{1cm}\hdashrule[3pt]{1cm}{0.5pt}{1mm} & \hspace{1cm}\hdashrule[3pt]{1cm}{0.5pt}{1mm} & $\mathfrak{B}_D \overline{T}=i\overline{D}$
    \end{tabular}}
    \caption{ BRST transformations of the Hodge duals. $N_{ghost}$ stands for the ghost number.}
\end{table}

\subsection{Gauge-fixing the  Weyl invariance and the complete BRST operator.}

Let us recall that in our formalism, in the path integral defining our theory, we must integrate over the unconstrained field $g_{\mu\nu}$, not over $\hat{g}_{\mu\nu}$. Also recall $\hat{g}_{\mu\nu}$ is defined in terms of $g_{\mu\nu}$ by Eq.~(\ref{unconstrainedg}). Now, it is the unimodular metric $\hat{g}_{\mu\nu}$, the metric that enters $\mathcal{L}_\text{ghost}^{(D)}$ and, hence, $\mathcal{L}_\text{ghost}^{(D)}$  depends on $g_{\mu\nu}$ only through $\hat{g}_{\mu\nu}$. It then follows that $\mathcal{L}_\text{ghost}^{(D)}$  has a gauge symmetry, the following Weyl symmetry:
\begin{equation}
g_{\mu\nu}(x)\to g_{\mu\nu}(x)+\delta_{\text{W}}g_{\mu\nu}(x),\quad \delta_{\text{W}}g_{\mu\nu}(x)=2\theta(x)g_{\mu\nu}(x),
\label{WeylT}
\end{equation}
where $\theta$ is an arbitrary scalar field.

The Weyl symmetry we have just mentioned has to be removed from the action in the path integral, lest the path integral be ill defined. We shall do so by using the BRST formalism.

Since one of the purposes of this paper is to make consistent the formalism of Ref.~\cite{Kugo:2022iob} with the classic linear splitting of the gravitational field,  we shall introduce that splitting now:
\begin{equation}
g_{\mu\nu}=\overline{g}_{\mu\nu}+h_{\mu\nu},
\label{linearsplitting}
\end{equation}
and define the BRST transformations so that they are adapted to the splitting in question. In the previous equation $\overline{g}_{\mu\nu}$ is the background gravitational field and $h_{\mu\nu}$ is the quantum gravitational field.

Now, let $R$ ($N_\text{ghost}(R)=1$) denote the ghost associated to the Weyl transformation in Eq.~(\ref{WeylT}),  $\overline{R}$ ($N_\text{ghost}(\overline{R})=-1$) the corresponding antighost field and $L$ ($N_\text{ghost}(L)=0$) the \textit{Nakanishi-Lautrup} auxiliary field. Let $\mathfrak{B}_W$ denote the BRST operator for the Weyl symmetry in Eq.~(\ref{WeylT}). Then, we define the action of $\mathfrak{B}_W$ on the fields as follows

\begin{equation}
\begin{array}{l}
     {\mathfrak{B}_W \overline{g}_{\mu\nu}=0,\quad\mathfrak{B}_W h_{\mu\nu}=2R\left(\bar{g}_{\mu\nu}+h_{\mu\nu}\right),}\\[4pt]
      {\mathfrak{B}_W R=0,}\\[4pt]
    {\mathfrak{B}_W \overline{R}=L,\quad\mathfrak{B}_W L=0.}
\end{array}
\label{WeylBRST}
\end{equation}

We will also need  the action of $\mathfrak{B}_W$ on the first generation, the second generation, the third generation ghosts and the corresponding \textit{Nakanishi-Lautrup} auxiliary fields  introduced in the previous subsections. We define  this action to be equal to zero. Hence, we have
\begin{equation}
\begin{array}{l}
{\mathfrak{B}_W B_{\mu\nu}=0,\quad\mathfrak{B}_W C _{\mu\nu}=0,\quad\mathfrak{B}_W \overline{C}_{\mu\nu}=0,\quad\mathfrak{B}_W D_{\mu}=0,\quad
\mathfrak{B}_W \overline{D}_{\mu}=0,}\\[4pt]
{\mathfrak{B}_W T_{\mu\nu}=0,\quad\mathfrak{B}_W \overline{T}_{\mu\nu}=0,\quad\mathfrak{B}_W D=0,\quad
\mathfrak{B}_W \overline{D}=0,}
\end{array}
\label{WeylBRSTforgenerations}
\end{equation}
which are the fields entering the ghost Lagrangian in Eq.~(\ref{translagrangian}), in addition to the unimodular metric $\hat{g}_{\mu\nu}$.

Obviously, $\mathfrak{B}_W^2=0$ --recall that $R$ is a Grassmann field. Eqs.~(\ref{unconstrainedg}) and (\ref{WeylBRST}) imply
\begin{equation}
\mathfrak{B}_W \hat{g}_{\mu\nu}=0.
\label{weylhatg}
\end{equation}

The fields $R$, $\overline{R}$ and $L$ are scalar fields and, hence, the action on these fields of the BRST operator, $\mathfrak{B}_D$, associated to the transverse diffeomorphisms reads
\begin{equation*}
\mathfrak{B}_D R=c_\text{T}^\rho\partial_\rho R,\quad\mathfrak{B}_D \overline{R}=c_\text{T}^\rho\partial_\rho\overline{R},\quad
\mathfrak{B}_D L=c_\text{T}^\rho\partial_\rho L,
\label{transRS}
\end{equation*}
where it is important to always bear in mind  that $c_\text{T}^\mu=-\varepsilon^{\mu\nu\rho\sigma}\hat{\nabla}_\nu C_{\rho\sigma}$.

Let us show now that $\mathfrak{B}_D^2=0$ on $R$, $\overline{R}$ and $L$. Let $F$ stand for any of those fields, then
\begin{equation*}
\begin{array}{l}
{\mathfrak{B}_D^2 F=\mathfrak{B}_D(c_\text{T}^\rho\partial_\rho F)=(\mathfrak{B}_D c_\text{T}^\rho)\partial_\rho F-c_\text{T}^\rho\partial_\rho \mathfrak{B}_D F = c_\text{T}^\nu\partial_\nu c_\text{T}^\rho\partial_\rho F-c_\text{T}^\nu\partial_\nu( c_\text{T}^\rho\partial_\rho F)=0.}
\end{array}
\end{equation*}
We have taken into account that $c_\text{T}^\rho$ is a Grassmann field and that Eqs.~(\ref{BcT}) and (\ref{usefuleq}) hold.

Next, let us define the following linear operator
\begin{equation}
\mathfrak{B}=\mathfrak{B}_D+\mathfrak{B}_W
\label{completeBRSTop}
\end{equation}
acting on the fields we have introduced so far.

Let us now prove that $\mathfrak{B}$ is a BRST operator, i.e., that $\mathfrak{B}^2=0$. Since we already know that $\mathfrak{B}_D^2=0$ and $\mathfrak{B}_W^2=0$, all that is left for us to do is show that
\begin{equation}
\mathfrak{B}_D \mathfrak{B}_W+\mathfrak{B}_W \mathfrak{B}_D=0.
\label{compatible}
\end{equation}

Taking into account Eqs.~(\ref{BRSTforhmn}) and (\ref{WeylBRST}), one gets
\begin{eqnarray*}
    \mathfrak{B}_D\mathfrak{B}_W h_{\mu\nu}&=&2g_{\mu\nu}c_\text{T}^\rho\partial_\rho R-2\mathfrak{B}_Dh_{\mu\nu},\nonumber\\
    \mathfrak{B}_W\mathfrak{B}_D h_{\mu\nu}&=&-2g_{\mu\nu}c_\text{T}^\rho\partial_\rho R+2\mathfrak{B}_Dh_{\mu\nu}.
\end{eqnarray*}
\begin{eqnarray*}
    \mathfrak{B}_D\mathfrak{B}_W \overline{R}&=&\mathfrak{B}_D L=c_\text{T}^\rho\partial_\rho L,\nonumber\\
    \mathfrak{B}_W\mathfrak{B}_D \overline{R}&=&\mathfrak{B}_W\left(c_\text{T}^\rho\partial_\rho \overline{R}\right)=-c_\text{T}^\rho\partial_\rho L.
\end{eqnarray*}

That $\mathfrak{B}_W c_\text{T}^\mu=0$ leads immediately to the conclusion that Eq.~(\ref{compatible}) holds when acting on $R$ and $L$. Now, recall that $\mathfrak{B}_W$ annihilates the first generation, the second generation, the third generation ghosts and the corresponding \textit{Nakanishi-Lautrup} auxiliary fields, and that $\mathfrak{B}_D$ acting on those very fields \textit{only} involves the fields in question and the unimodular metric $\hat{g}_{\mu\nu}$ --this metric being invariant under the Weyl transformations of $g_{\mu\nu}.$ Hence, Eq.~(\ref{compatible}) also holds for the fields we have just mentioned.

We are now ready to carry out the BRST gauge-fixing of the Weyl gauge symmetry in Eq.~(\ref{WeylT}) in a way which is consistent with the BRST gauge-fixing of the transverse diffeomorphisms. Indeed, we shall add to $\mathcal{L}_\text{ghost}^{(D)}$ in Eq.~(\ref{translagrangian}) --which does the gauge-fixing of the transverse diffeomorphism invariance-- the following term
\begin{equation}
\begin{array}{l}
   { \mathcal{L}_\text{ghost}^{(W)}=\mathfrak{B}\big[\overline{\nabla}_{\mu} \overline{R} \cdot\overline{\nabla}^{\mu}\big(L-\alpha' h \big)\big]=
   \mathfrak{B}_D\big[\overline{\nabla}_{\mu} \overline{R} \cdot\overline{\nabla}^{\mu}\big(L-\alpha' h \big)\big]+
   \mathfrak{B}_W\big[\overline{\nabla}_{\mu} \overline{R} \cdot\overline{\nabla}^{\mu}\big(L-\alpha' h \big)\big]=}\\[4pt]
   {\quad\quad\overline{\nabla}_{\mu}
   ig\big[c_\text{T}^\rho\overline{\nabla}_\rho \overline{R}\big] \cdot\overline{\nabla}^{\mu}\big(L-\alpha' h \big)
    - \overline{\nabla}_{\mu} \overline{R} \cdot\overline{\nabla}^{\mu}\big[c_\text{T}^\rho\overline{\nabla}_\rho L\big]+\alpha'\overline{\nabla}_{\mu} \overline{R} \cdot
    \overline{\nabla}^{\mu}\big[2\overline{\nabla}^\rho c^\text{T}_\rho+c_\text{T}^\rho\overline{\nabla}_\rho h}\\[4pt]
    {\quad\quad+2\overline{\nabla}^\sigma c_\text{T}^\rho h_{\rho\sigma}\big]+
\overline{\nabla}_{\mu} L \cdot\overline{\nabla}^{\mu}\big(L-\alpha' h \big) + 2 \alpha' \big(4+h\big)\overline{\nabla}_{\mu} \overline{R} \cdot \overline{\nabla}^{\mu} R
    +2 \alpha' \overline{\nabla}_{\mu} \overline{R} \cdot R\overline{\nabla}^{\mu} h},
\end{array}
\label{Weyllagrangian}
\end{equation}
where $h=\overline{g}^{\mu\nu}h_{\mu\nu}$ and $\overline{\nabla}_\mu$ is the covariant derivative with respect to $\overline{g}_{\mu\nu}$.

Notice that $\mathcal{L}_\text{ghost}^{(W)}$ is exact with regard to the complete BRST operator $\mathfrak{B}$ in Eq.~(\ref{completeBRSTop}), a must if the full power of the BRST formalism is to be taken advantage of. But what about $\mathcal{L}_\text{ghost}^{(D)}$ in Eq.~(\ref{translagrangian})? The answer is in the affirmative. Indeed, $\mathcal{L}_\text{ghost,1}^{(D)}$ in Eq, (\ref{Lagrangian1}), $\mathcal{L}_\text{ghost,2}^{(D)}$ in Eq, (\ref{Lagrangian2}) and $\mathcal{L}_\text{ghost,3}^{(D)}$ in Eq, (\ref{Lagrangian3}) can be also expressed as follows
\begin{equation*}
\begin{array}{l}
{\mathcal{L}_\text{ghost,1}^{(D)}=
    -i\mathfrak{B}\left[\hat{g}_{\mu\nu}\hat{\nabla}_\rho\overline{c}^{\rho\nu}\cdot\partial_\lambda \hat{g}^{\lambda\mu}\right]}\\[4pt]
{\mathcal{L}_\text{ghost,2}^{(D)}=
\frac{i}{2}\mathfrak{B}\left[\overline{d}^{\rho\nu\mu}\hat{\nabla}_\rho c_{\nu\mu}+b^{\rho\nu\mu}\hat{\nabla}_\rho \overline{c}_{\nu\mu}\right]}\\[4pt]
{\mathcal{L}_\text{ghost,3}^{(D)}=
\frac{-i}{6}\mathfrak{B}\left[-\overline{t}^{\sigma\rho\nu\mu}\left(\hat{\nabla}_\sigma d_{\rho\nu\mu}+\frac{\alpha}{4}d_{\sigma\rho\nu\mu}\right)+c^{\sigma\rho\nu\mu}\hat{\nabla}_\sigma \overline{d}_{\rho\nu\mu}+\overline{c}^{\sigma\rho\nu\mu}\hat{\nabla}_\sigma b_{\rho\nu\mu}\right],}
\end{array}
\end{equation*}
This is a consequence of the fact that, as we have seen above, the action of $\mathfrak{B}_W$ on any of the fields entering the definition of
$\mathcal{L}_\text{ghost,1}^{(D)}$, $\mathcal{L}_\text{ghost,2}^{(D)}$ and $\mathcal{L}_\text{ghost,3}^{(D)}$ yields zero. We thus conclude that fixing the gauge for the Weyl symmetry does not introduce any modification  of $\mathcal{L}_\text{ghost}^{(D)}$ as given in Eqs.~(\ref{translagrangian}), (\ref{Lagra1}), (\ref{Lagra2}) and (\ref{Lagra3}).

\subsection{The complete BRST invariant action.}

In the previous subsections we have achieved, by using the BRST formalism, a complete gauge-fixing of the two gauge symmetries of our theory.  Let us now quote  the full action, $\mathcal{S}$, of our theory. This action reads
\begin{equation}
\mathcal{S}=\mathcal{S}_{class}+2\int d^4x\, \mathcal{L}_\text{ghost}^{(D)}+2\int d^4x\,\mathcal{L}_\text{ghost}^{(W)}.
\label{fullaction}
\end{equation}
$\mathcal{S}_{class}$, $\mathcal{L}_\text{ghost}^{(D)}$ and $\mathcal{L}_\text{ghost}^{(W)}$ are given in Eqs.~(\ref{clasaction}), (\ref{translagrangian}) and (\ref{Weyllagrangian}), respectively.

\section{The free graviton propagator and Newton's Law of gravitation.}

In this section we shall put to work the action $\mathcal{S}$, in Eq.~(\ref{fullaction}), and work out the free graviton propagator, first, and, then, obtain Newton's law of gravitation by using Feynman diagrams.

Let the background field $\overline{g}_{\mu\nu}$ in Eq.~(\ref{linearsplitting}) be the Minkowski metric with signature $(-,+,+,+)$. Then, the free part of
$\mathcal{S}$ needed for the computation of the free graviton propagator reads
\begin{equation}
\begin{array}{l}
    {\int d^4x\,\Big\{\frac{1}{4} h^{\mu\nu}\Box h_{\mu\nu} +\frac{1}{2}\left(\partial_\mu h^{\mu\nu}\right)^2 + \frac{1}{4}h_{\mu\nu}\partial^\mu\partial^\nu h- \frac{1}{32} h \Box h + \frac{1}{2} \varepsilon_{\mu\nu\rho\sigma}\partial^\nu B^{\rho\sigma}\cdot\partial_\lambda h^{\lambda\mu}+B_\mu \partial_\nu B^{\nu\mu}}\\[4pt]
    {+B\partial^\mu B_\mu+\alpha' L \Box h -L \Box L
    -\overline{D}^{\mu}\left(\Box D_{\mu}-\partial_\mu\partial^\nu D_\nu\right)+\overline{D}\partial^\mu D_{\mu} + \partial_\mu \overline{D}^{\mu}\cdot D +\alpha \overline{D} D\Big\}.}
 \end{array}
 \end{equation}
Hence, the free graviton propagator is obtained by inverting the following matrix
\begin{equation*}
\begin{matrix}
&& h_{\rho\sigma} & B_{\rho\sigma} & B_\rho & B & L &\\
&&  & \vspace{-2mm} &  &  &  &\\
h_{\mu\nu} &&
\mathcal{V}_{h,h}
& 2\varepsilon^{\kappa\lambda\rho\sigma}p_\lambda \delta^{(\nu}_\kappa p^{\mu)} & 0 & 0 & -4\alpha' p^2 \eta^{\mu\nu} &\\
B_{\mu\nu} && -2\varepsilon^{\mu\nu\kappa(\rho}p^{\sigma)}p_\kappa & 0 & i4\eta^{\rho[\mu}p^{\nu]} & 0 & 0 &\\
B_\mu &
\smash{\left(\begin{array}{c}
     \\  \\ \\ \\ \\ \\
\end{array}\right.\hspace{-4mm}}
& 0 & -i4\eta^{\mu[\rho}p^{\sigma]} & 0 & -i4p^\mu & 0 &
\smash{\hspace{-3mm}\left.\begin{array}{c}
     \\  \\ \\ \\ \\ \\
\end{array}\right)}\\
B && 0 & 0 & i4p^\rho & 0 & 0 &\\
L && -4\alpha' p^2\eta^{\rho\sigma} & 0 & 0 & 0 & 4p^2 &
\end{matrix},
\end{equation*}
where $\mathcal{V}_{h,h}=-p^2\eta^{\mu(\rho}\eta^{\sigma)\nu}+2p^{(\mu}\eta^{\nu)(\sigma}p^{\rho)}-\frac{1}{2}\left(p^\mu p^\nu\eta^{\rho\sigma}+p^\rho p^\sigma\eta^{\mu\nu}\right)-\frac{1}{8}p^2\eta^{\mu\nu}\eta^{\rho\sigma}$. For the free graviton propagator of our theory, we have obtained the following result:
\begin{equation}
\begin{array}{l}
{\langle h_{\mu_1\mu_2}(p) h_{\mu_3\mu_4}(-p)\rangle=}\\[4pt]
{\frac{i}{p^2}\Big\{A_1\,(\eta_{\mu_1\mu_2}\eta_{\nu_1\nu_2}+\eta_{\mu_1\nu_2}\eta_{\nu_1\mu_2})+A_2\eta_{\mu_1\nu_1}\eta_{\mu_2\nu_2}
+A_3\,\frac{1}{p^2}(\eta_{\mu_1\nu_1}p_{\mu_2}p_{\nu_2}+\eta_{\mu_2\nu_2}p_{\mu_1}p_{\nu_1})}\\[4pt]
{+A_4\,\frac{1}{p^2}(\eta_{\mu_1\mu_2}p_{
  \nu_1}p_{\nu_2}+\eta_{\mu_1\nu_2}p_{\nu_1}p_{\mu_2}+\eta_{\nu_1\mu_2}p_{\mu_1}p_{\nu_2}+\eta_{\nu_1\nu_2}p_{\mu_1}p_{\mu_2})+A_5\,
  \frac{1}{p^4}p_{\mu_1}p_{\nu_1}p_{\mu_2}p_{\nu_2}\Big\},}
  \end{array}
  \label{gravitonprop}
  \end{equation}
with
\begin{equation*}
A_1=-\frac{1}{2},\quad  A_2= \frac{3+32{\alpha'}^2}{8(1+8{\alpha'}^2)},\quad A_3=-1,\quad
    A_4=\frac{1}{2},\quad A_5=2.
\end{equation*}

In unimodular gravity, the graviton field $h_{\mu\nu}$ couples to the traceless part of the energy-momentum tensor $\hat{T}^{\mu\nu}\equiv T^{\mu\nu}-\eta^{\mu\nu}\, T^\rho_\rho/4$. This is equivalent to saying that it is the traceless part of graviton field,  $\hat{h}_{\mu\nu}= h_{\mu\nu}-\eta_{\mu\nu} h/4$, the one which couples to the energy-momentum tensor $T^{\mu\nu}$. Hence, to retrieve the gravitational Newton's potential from unimodular gravity it is useful  to compute the
free two-point Green function of $\hat{h}_{\mu\nu}$. This Green function can be obtained from Eq.~(\ref{gravitonprop}) and it reads
\begin{equation}
\begin{array}{l}
{\langle\hat{h}_{\mu_1\nu_1}(p)\hat{h}_{\mu_2\nu_2}(-p)\rangle =}\\[4pt] {\frac{i}{p^2}\Big\{-\frac{1}{2}\,(\eta_{\mu_1\mu_2}\eta_{\nu_1\nu_2}+\eta_{\mu_1\nu_2}\eta_{\nu_1\mu_2}-\eta_{\mu_1\nu_1}\eta_{\mu_2\nu_2})
-\,\frac{1}{p^2}(\eta_{\mu_1\nu_1}p_{\mu_2}p_{\nu_2}+\eta_{\mu_2\nu_2}p_{\mu_1}p_{\nu_1}}\\[4pt]
{+\frac{1}{2}\,\frac{1}{p^2}(\eta_{\mu_1\mu_2}p_{\nu_1}p_{\nu_2}+\eta_{\mu_1\nu_2}p_{\nu_1}p_{\mu_2}+\eta_{\nu_1\mu_2}p_{\mu_1}p_{\nu_2}+
\eta_{\nu_1\nu_2}p_{\mu_1}p_{\mu_2})+2\,\frac{1}{p^4}p_{\mu_1}p_{\nu_1}p_{\mu_2}p_{\nu_2}\Big\}.}
\end{array}
\label{tracelessprop}
\end{equation}

The Gravitational Newton's potential can be obtained by computing, in the non-relativistic limit, the amplitude, $\mathcal{M}$, of the process in which one graviton  is exchanged between two scalar particles with masses $M_1$ y $M_2$. The Feynman diagram describing this process is in Fig.~1. The amplitude in question is given by
\begin{equation}
    \mathcal{M}=-\frac{i}{4} T^{(1)}_{\mu\nu}(p_1,p'_1)\langle\hat{h}^{\mu\nu}(k)\hat{h}^{\rho\sigma}(-k)\rangle T^{(2)}_{\rho\sigma}(p_2,p'_2),
\label{amplitude}
\end{equation}
where $p_i$ y $p'_i$  are the initial and final momenta, respectively, of the $i$ particle, $i=1,2$. Hence, the transferred momentum $k$ is given by $k=p_1-p'_1=p'_2-p_2$ y $M_i^2=-p_i^2=-{p'}_i^2$. In Eq.~(\ref{amplitude}) $T^{(i)}_{\mu\nu}(p_i,p'_i)$ denotes the on-shell lowest order contribution of the energy-momentum tensor between the states with momenta $p_i$ y $p'_i$. $T^{(i)}_{\mu\nu}(p_i,p'_i)$ is given by
\begin{equation*}
    T^{(i)}_{\mu\nu}(p_i,p'_i)\equiv\langle p'_i|T_{\mu\nu}|p_i\rangle={p_i}_\mu{p'_i}_\nu+{p_i}_\nu{p'_i}_\mu+\frac{1}{2}k^2\eta_{\mu\nu},
\end{equation*}
which in the non-relativistic limit --hence $k^\mu=(0,\vec{k})$-- reads
\begin{equation*}
    T^{(i)}_{\mu\nu}(p_i,p'_i)=-2M_i^2\eta_{\mu 0}\eta_{\nu 0}.
\end{equation*}
We thus conclude that, in the non-relativistic, limit the amplitude $\mathcal{M}$ is equal to
\begin{equation*}
\mathcal{M}^{nonrel}=-i M_1^2 M_2^2\langle\hat{h}_{00}(k)\hat{h}_{00}(-k)\rangle,
\end{equation*}
where $\langle\hat{h}_{00}(k)\hat{h}_{00}(-k)\rangle$ is given in Eq.~(\ref{tracelessprop}) and $k^\mu=(0,\vec{k})$.

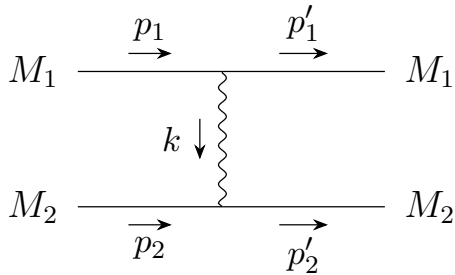
\begin{figure}[ht]
\noindent\begin{minipage}[t]{\columnwidth}
\centering
\scalebox{1.2}{
\begin{tikzpicture}
  \begin{feynman}
    \vertex (a) {\(M_1\)};
    \vertex [right=2.1cm of a] (b);
    \vertex [right=1.8cm of b] (c) {\(M_1\)};
    \vertex [below=of b] (e);
    \vertex [left=2.1cm of e, below=of a] (d) {\(M_2\)};
    \vertex [right=1.8cm of e, below=of c] (f) {\(M_2\)};

    \diagram* {
      (a) -- [reversed momentum'={[arrow shorten=0.65,arrow distance=2mm]\small \(p_1\)}] (b) -- [reversed momentum'={[arrow shorten=0.65,arrow distance=2mm]\small \(p'_1\)}] (c),
      (d) -- [momentum'={[arrow shorten=0.35,arrow distance=2mm]\small \(p_2\)}] (e) -- [momentum'={[arrow shorten=0.35,arrow distance=2mm]\small \(p'_2\)}] (f),
      (b) -- [boson, momentum'={[arrow shorten=0.35,arrow distance=2.5mm]\small \(k\)}] (e),
    };
  \end{feynman}
\end{tikzpicture}}
\end{minipage}
    \caption{The exchange of one graviton}
    \label{fig:diagrama}
\end{figure}

Now, the Fourier transform of the Newton's potential is given by
\begin{equation*}
    \tilde{V}(\vec{k})=\frac{\mathcal{M}^{nonrel}}{2M_1 2M_2}=-\frac{i}{4} M_1 M_2\langle\hat{h}_{00}(k)\hat{h}_{00}(-k)\rangle.
\end{equation*}

From  Eq.~(\ref{tracelessprop}), we get $\langle\hat{h}_{00}(k)\hat{h}_{00}(-k)\rangle=-i/2k^2$ so that
\begin{equation*}
    \tilde{V}(\Vec{k})=-\frac{1}{8} \frac{M_1 M_2}{\Vec{k}^2},
\end{equation*}
which precisely the Fourier transform of the Newton's potential one finds in elementary textbooks, in the unit system where $G=1/(32\pi)$.

\section{The Slavnov-Taylor equation and the linearized Slavnov-Taylor operator.}

As was discussed by the authors of Ref.~\cite{Kugo:2022iob} the construction of their BRST formalism involving antisymmetric tensor ghosts is such that the field content of the resulting BRST invariant theory depends on the dimension of spacetime. We have also analyzed this issue right below Eq.~(\ref{newgaugesymone}). It is then clear that dimensional regularization does not provide a BRST invariant regularization of the  field theory in 4 dimensional spacetime --nor in any other spacetime dimension. Since no  regularization method that preserves general covariance, while keeping the spacetime dimension equal to 4, is known, one is lead to the conclusion that one has to take advantage of the algebraic renormalization techniques --see Ref.~\cite{Piguet:1995er}-- to keep the BRST symmetry of the renormalized theory under control. The implementation of algebraic renormalization techniques begins with the derivation of the formal Slavnov-Taylor equation, which is to be satisfied by the 1PI functional, and continues with the formulation, and solution, of a cohomology problem for the linearized Slavnov-Taylor operator. In this section,  we will just obtain the Slavnov-Taylor equation for the 1PI functional and define the linearized Slavnov-Taylor  and its cohomology problem. Since the theory is not renormalizable by power counting, the solution of this cohomology problem will demand the use of  the highbrow algebraic cohomological and homological methods reviewed in Ref.~\cite{Barnich:2000zw} and this lies quite outside the scope of our paper.

Let us begin the derivation of the formal Slavnov-Taylor equation. Let $\mathcal{S}$ be the action in Eq.~(\ref{fullaction}). Let $\Phi$ generically denote the fields of our theory. These fields are
\begin{equation*}
h_{\mu\nu},B_{\mu\nu}, B_{\mu}, B, C_{\mu\nu}, C_{\mu}, C,\overline{C}_{\mu\nu}, \overline{C}_{\mu}, \overline{C}, D_{\mu}, D, \overline{D}_{\mu}, \overline{D}, T, \overline{T}, R,  \overline{R}\;\text{and}\; L.
\end{equation*}
Then, the generating functional of the complete Green functions of our theory reads
\begin{equation}
    Z\left[J^{\mu\nu},j^{\mu\nu},...\right]=\int\mathcal{D}\Phi\, \ \left[\mathcal{D}\text{ghost}\right]\; e^{i\mathcal{S}+ i\int d^4x\, \mathcal{C}},
\label{partitionfunction}
\end{equation}
where
\begin{equation*}
\begin{array}{l}
    {\mathcal{C}=J^{\mu\nu}h_{\mu\nu}+j^{\mu\nu}B_{\mu\nu}+ j^\mu B_{\mu} + j B + fL
     +\overline{\omega}^{\mu\nu} C_{\mu\nu} +\overline{\omega}^{\mu}C_{\mu} +\overline{\omega} C +\overline{C}_{\mu\nu}\omega^{\mu\nu}}\\[4pt]
    {\phantom{\mathcal{C}=} +\overline{C}_{\mu}\omega^{\mu} +\overline{C}\omega + \overline{q} R + \overline{R} q +\overline{k}^\mu D_{\mu} +\overline{k} D
     +\overline{D}_{\mu} k^\mu +\overline{D} k +\overline{\chi} T +\overline{T} \chi +\xi^{\mu\nu}\mathfrak{B}h_{\mu\nu}}\\[4pt]
    {\phantom{\mathcal{C}=} +\phi \mathfrak{B} L +\rho^{\mu\nu} \mathfrak{B} C_{\mu\nu} + \tau \mathfrak{B} R + \sigma \mathfrak{B} \overline{R}
    +\rho^\mu \mathfrak{B} D_{\mu} +\rho \mathfrak{B} T.}
\end{array}
\end{equation*}
Recall that $\mathfrak{B}$ is the complete BRST operator in Eq.~(\ref{completeBRSTop}). Notice that we have included external fields coupled to the BRST variations
\begin{equation*}
\mathfrak{B}h_{\mu\nu},\, \mathfrak{B} L,\, \mathfrak{B} C_{\mu\nu},\mathfrak{B} R,\, \mathfrak{B} \overline{R},\, \mathfrak{B} D_{\mu}\;\text{and}\; \mathfrak{B} T,
\end{equation*}
for those variations are composite operators.

Taking into account that, by definition, the external fields $J^{\mu\nu}$, $j^{\mu\nu}$,.... are annihilated by $\mathfrak{B}$ and so it is $\mathcal{S}$, we conclude that
\begin{equation}
\begin{array}{l}
    {\langle \int d^4x \big(J^{\mu\nu}\mathfrak{B} h_{\mu\nu}+ j^\mu C_{\mu} + f \mathfrak{B} L -\overline{\omega}^{\mu\nu} \mathfrak{B} C_{\mu\nu}
    -i\overline{\omega}D +iB_{\mu\nu}\omega^{\mu\nu}
    +iB\omega }\\[4pt]
    {\phantom{\langle \int d^4x \big(J^{\mu\nu}\mathfrak{B} h_{\mu\nu}+ j^\mu C_{\mu} + f \mathfrak{B} L }
    -\overline{q} \mathfrak{B} R+q \mathfrak{B}\overline{R}
    +\overline{k}^\mu \mathfrak{B} D_{\mu} +\overline{C}_{\mu}k^\mu -\overline{\chi} \mathfrak{B} T +i\overline{D} \chi\big)\rangle =0.}
    \end{array}
    \label{meanvalueeq}
\end{equation}
where $\langle A \rangle\equiv \int\mathcal{D}\Phi \ A \cdot \text{exp}\left[{i\mathcal{S}+i\mathcal{C})}\right]$. The previous equation is obtained by applying $\mathfrak{B}$ to both sides of Eq.~(\ref{partitionfunction}).

Eq.~(\ref{meanvalueeq}) can be recast into the following form
\begin{equation}
\begin{array}{l}
    {\Big[\int d^4x \big(J^{\mu\nu}\frac{\delta}{\delta \xi^{\mu\nu}}+ j^\mu \frac{\delta}{\delta \overline{\omega}^{\mu}} + f \frac{\delta}{\delta \phi} -\overline{\omega}^{\mu\nu} \frac{\delta}{\delta \rho^{\mu\nu}}
    -i\overline{\omega}\frac{\delta}{\delta \overline{k}} +i\omega^{\mu\nu}\frac{\delta}{\delta j^{\mu\nu}}}\\[4pt]
    {\quad\quad\quad\quad+i\omega \frac{\delta}{\delta j} -\overline{q} \frac{\delta}{\delta \tau}+q \frac{\delta}{\delta \sigma}
    +\overline{k}^\mu \frac{\delta}{\delta \rho^{\mu}} -k^\mu \frac{\delta}{\delta \omega^{\mu}}-\overline{\chi}
    \frac{\delta}{\delta \rho} +i\chi\frac{\delta}{\delta k} \big)\Big]Z=0.}
\end{array}
\label{Zeq}
\end{equation}

Let $\mathfrak{J}$ be given by
\begin{equation*}
\begin{array}{l}
    {\mathfrak{J}=J^{\mu\nu}h_{\mu\nu}+j^{\mu\nu}B_{\mu\nu}+ j^\mu B_{\mu} + j B + fL
     +\overline{\omega}^{\mu\nu} C_{\mu\nu} +\overline{\omega}^{\mu}C_{\mu} +\overline{\omega} C +\overline{C}_{\mu\nu}\omega^{\mu\nu}}\\[4pt]
    {\phantom{\mathcal{C}=} +\overline{C}_{\mu}\omega^{\mu} +\overline{C}\omega + \overline{q} R + \overline{R} q +\overline{k}^\mu D_{\mu} +\overline{k} D
     +\overline{D}_{\mu} k^\mu +\overline{D} k +\overline{\chi} T +\overline{T} \chi .}
     \end{array}
\end{equation*}
Then, the 1PI functional of our theory is defined as usual:
\begin{equation*}
    \Gamma\left[\Phi\right]=W\left[J^{\mu\nu},j^{\mu\nu},...\right]-\int d^4x\ \mathfrak{J},
\end{equation*}
where $W\left[J^{\mu\nu},j^{\mu\nu},...\right]$ is such that
\begin{equation*}
Z\left[J^{\mu\nu},j^{\mu\nu},...\right]/Z[0]=\text{exp}\left(i W\left[J^{\mu\nu},j^{\mu\nu},...\right]\right)
\end{equation*}
Using these definitions is not difficult to show that Eq.~(\ref{Zeq}) is equivalent to the following equation
\begin{equation*}
\begin{array}{l}
    {\int d^4x \Big[\dfrac{\delta \Gamma}{\delta h_{\mu\nu}}\dfrac{\delta\Gamma}{\delta \xi^{\mu\nu}}+ \dfrac{\delta \Gamma}{\delta B_{\mu}} C_\mu + \dfrac{\delta \Gamma}{\delta L} \dfrac{\delta\Gamma}{\delta \phi} + \dfrac{\delta \Gamma}{\delta C_{\mu\nu}} \dfrac{\delta\Gamma}{\delta \rho^{\mu\nu}}
    +i\dfrac{\delta \Gamma}{\delta C}D +i\dfrac{\delta \Gamma}{\delta \overline{C}_{\mu\nu}}B_{\mu\nu}}\\[4pt]
    {\phantom{\int d^4x \Big[}
    +i\dfrac{\delta \Gamma}{\delta \overline{C}} B +\dfrac{\delta \Gamma}{\delta R} \dfrac{\delta\Gamma}{\delta \tau}+\dfrac{\delta \Gamma}{\delta \overline{R}} \dfrac{\delta\Gamma}{\delta \sigma}
    +\dfrac{\delta \Gamma}{\delta D_{\mu}} \dfrac{\delta\Gamma}{\delta \rho^{\mu}} +\dfrac{\delta \Gamma}{\delta \overline{D}_{\mu}} \overline{C}_\mu +\dfrac{\delta \Gamma}{\delta T} \dfrac{\delta\Gamma}{\delta \rho} +i\dfrac{\delta \Gamma}{\delta \overline{T}}\overline{D}\Big]=0.}
\end{array}
\end{equation*}
This is the Slavnov-Taylor equation. This equation governs the BRST equation at the quantum level.

By expanding $\Gamma\left[\Phi\right]$ in powers of $\hbar$, one reaches the conclusion that the Slavnov-Taylor equation implies that the one-loop contribution, $\Gamma_{1}\left[\Phi\right]$, to $\Gamma\left[\Phi\right]$ must satisfy the linearized Slavnov-Taylor equation, which reads
\begin{equation}
\Delta \Gamma_{1}\left[\Phi\right]=0,
\label{linSTeq}
\end{equation}
where $\Delta$ is the so-called linearized Slavnov-Taylor operator:
\begin{equation*}
\begin{array}{l}
{\Delta=}\\[4pt]
{\int d^4x \Big[\dfrac{\delta \Gamma_0}{\delta h_{\mu\nu}}\dfrac{\delta\phantom{\Gamma}}{\delta \xi^{\mu\nu}}+
                 \dfrac{\delta \Gamma_0}{\delta \xi_{\mu\nu}}\dfrac{\delta\phantom{\Gamma}}{\delta h^{\mu\nu}}+
C_\mu\dfrac{\delta\phantom{ \Gamma}}{\delta B_{\mu}} +
\dfrac{\delta \Gamma_0}{\delta L} \dfrac{\delta\phantom{\Gamma}}{\delta \phi}+
\dfrac{\delta \Gamma_0}{\delta \phi} \dfrac{\delta\phantom{\Gamma}}{\delta L}
+ \dfrac{\delta \Gamma_0}{\delta C_{\mu\nu}} \dfrac{\delta\phantom{\Gamma}}{\delta \rho^{\mu\nu}}+
\dfrac{\delta \Gamma_0}{\delta \rho_{\mu\nu}} \dfrac{\delta\phantom{\Gamma}}{\delta C^{\mu\nu}} }\\[4pt]
    {\phantom{\int d^4x \Big[}
    +iD\dfrac{\delta \phantom{\Gamma}}{\delta C} +iB_{\mu\nu}\dfrac{\delta \phantom{\Gamma}}{\delta \overline{C}_{\mu\nu}}
    +iB\dfrac{\delta \phantom{\Gamma}}{\delta \overline{C}}
    +\dfrac{\delta \Gamma_0}{\delta R} \dfrac{\delta\phantom{\Gamma}}{\delta \tau}
    +\dfrac{\delta \Gamma_0}{\delta \tau} \dfrac{\delta\phantom{\Gamma}}{\delta R}
    +\dfrac{\delta \Gamma_0}{\delta \overline{R}} \dfrac{\delta\phantom{\Gamma}}{\delta \sigma}
    +\dfrac{\delta \Gamma_0}{\delta D_{\mu}} \dfrac{\delta\phantom{\Gamma}}{\delta \rho^{\mu}}
     +\dfrac{\delta \Gamma_0}{\delta \rho_{\mu}} \dfrac{\delta\phantom{\Gamma}}{\delta D^{\mu}} }\\[4pt]
     {\phantom{\int d^4x \Big[}
     +\overline{C}_\mu \dfrac{\delta \phantom{\Gamma}}{\delta \overline{D}_{\mu}}+
     \dfrac{\delta \Gamma_0}{\delta T} \dfrac{\delta\phantom{\Gamma}}{\delta \rho}+
     \dfrac{\delta \Gamma_0}{\delta \rho} \dfrac{\delta\phantom{\Gamma}}{\delta T}
     +i\overline{D}\dfrac{\delta \phantom{\Gamma}}{\delta \overline{T}}\Big],}
\end{array}
\end{equation*}
where
\begin{equation*}
\Gamma_{0}=\mathcal{S}+\int d^4x\,\big[\xi^{\mu\nu}\mathfrak{B}h_{\mu\nu}+\phi \mathfrak{B} L +\rho^{\mu\nu} \mathfrak{B} C_{\mu\nu} + \tau \mathfrak{B} R + \sigma \mathfrak{B} \overline{R}+\rho^\mu \mathfrak{B} D_{\mu} +\rho \mathfrak{B} T\big].
\end{equation*}
and $\mathcal{S}$ is given in Eq.~(\ref{fullaction}).

It can be shown that the linearized Slavnov-Taylor operator is nilpotent
\begin{equation*}
\Delta^2=0.
\end{equation*}
and it has ghost number equal to 1. Hence, any one-loop anomalous breaking, $\mathcal{A}$, of the linearized Slavnov-Taylor equation  --thus breaking the BRST symmetry-- must satisfy
\begin{equation*}
\Delta \mathcal{A} =0,\quad \mathcal{A}\neq \Delta \Xi,
\end{equation*}
$\mathcal{A}$ being a (infinite) linear  combination of monomials in the fields and their derivatives which has ghost number equal to 1. $\Xi$ is also a linear combination of monomial of the fields and their derivatives but it has got zero  ghost number. The solution to this problem is a very involved one and lies well beyond the scope of paper. If the BRST symmetry can be restored  at one-loop --i.e, $\mathcal{A}$=0, one has to face the very same type of cohomology problem at two loops, and so on.

\section{Conclusions.}

The main conclusion of the paper is that a  modification of the BRST transformations of Ref.~\cite{Kugo:2022iob} can be combined with the BRST transformations arising from the extra Weyl symmetry of our unimodular theory  into a set of transformations generated by a single BRST operator. In principle, that this   combination be feasible is a non trivial fact, for it involves different types of BRST transformations.  We obtained the full gauge-fixed action of the theory and use it to compute the Newton's potential and the Slavnov-Taylor equation, the latter governs full BRST symmetry of the theory at the quantum level. Let us point out that the derivation of the Slavnov-Taylor equation is a must. Indeed, there is no known regularization method that preserves the symmetries of the theory and therefore non BRST invariant counterterms must be introduced to compensate for the breaking of BRST invariance due to the regularization and/or renormalization process one uses. Recall that the ghost content of the theory depends on the spacetime dimension: dimensional regularization does not furnish a BRST invariant regularized theory. Also recall that the construction of those non BRST invariant counterterms and the analysis of their consistency is guided by the cohomology of the linearized Slavnov-Taylor operator.

\section{Acknowledgements.}
David Garc\'{\i}a L\'opez acknowledges financial support from the UCM-Instituto de F\'{\i}sica de Part\'{\i}culas y del Cosmos (IPARCOS).

\end{document}